\documentstyle[12pt,amssymb,amsfonts,amsbsy]{article}
\begin{document} 
\author{Ilja Schmelzer\thanks
       {WIAS Berlin}}

 \title{Quantization of Gravity Based on a Condensed Matter Model}

\begin{abstract}
\sloppypar

One way the ultraviolet problem may be solved is explicit physical
regularization. In this scenario, QFT is only the long distance limit
of some unknown non-Poincare-invariant microscopic theory.  One can
ask how complex and contrived such microscopic theories should be.

We show that condensed matter in standard Newtonian framework is
sufficient to obtain gravity.  We derive a metrical theory of gravity
with two additional to GR cosmological constants.  The observable
difference is similar to homogeneously distributed dark matter with $p
= -{1\over3}\varepsilon$ resp.  $p = \varepsilon$.  The gravitational
collapse stops before horizon formation and evaporates by Hawking
radiation.  The cutoff is not the Planck length, but expanding
together with the universe.  Thus, in some cosmological future
microscopic effects become observable.  

\end{abstract}

\maketitle

\section{Introduction}

Most of the research in quantum gravity relies on the assumption that
current observable symmetries, especially local Lorentz symmetry, are
fundamental symmetries of our universe.  Different directions (Grand
Unification, super-symmetry, tetrad formalism, Ashtekar variables,
strings, M-theory) may be interpreted as different attempts to
interpret current observable symmetries as parts of greater, more
fundamental symmetry groups.  A high degree of symmetry is required
to cure the problems with ultraviolet divergences.

But the observable symmetry groups may be simply a consequence of
``blindness for details'' at long distances.  In this case, below a
critical length QFT should be replaced by an unknown microscopic
theory which does not have these symmetries.  In this case, the
ultraviolet problems are cured by explicit, physical regularization.
Unlike in renormalized QFT, here the relationship between bare and
renormalized parameters obtains a physical meaning.  Such ideas are
quite old and in some aspects commonly accepted among particle
physicists \cite{Jegerlehner}.

Usually it is expected that the critical cutoff length is of order of
the Planck length $a_P \approx 10^{-33}cm$.  Below this distance, not
only Poincare invariance may break down. Even the laws of quantum
mechanics need not hold. Instead of a single space-time we may have to
face topological foam.  There seems to be no chance that such theories
lead to testable predictions.

The theory we present here shows that these expectations are not
justified.  First, the microscopic theory we present here is
surprisingly simple -- usual condensed matter in a Newtonian
framework, canonically quantized in classical Schr\"odinger theory.
Second, the cutoff length in this theory is not related with Planck
length.  In the expanding universe, it looks expanding together with
the universe.  Thus, in some cosmological future we will be able to
test the microscopic theory even with our current experimental
devices.  Last not least, the theory leads to testable predictions --
we obtain two additional cosmological constants.  This leads to
observable effects similar to homogeneously distributed dark matter
with $p = -1/3\varepsilon$ resp.  $p = \varepsilon$.

\section{The Theory}

The microscopic theory we consider is a standard condensed matter
theory in the Newtonian framework $R^3\otimes R$. That means, for long
distances it should be described by a density $\rho(x,t)$, a velocity
$v^a(x,t)$, and a stress tensor $\sigma^{ab}(x,t)$.  There may be some
additional inner steps of freedom $\phi^m(x,t)$, but no external
forces.  That's why the conservation laws have the following form:

\begin{eqnarray} \label{conservationlaw}
\partial_t \rho + \partial_i (\rho v^i) &= &0 \\
\partial_t (\rho v^j) + \partial_i(\rho v^i v^j - \sigma^{ij}) &= &0
\end{eqnarray}

Now we introduce new variables.  First, we combine the ten classical
steps of freedom into a metrical tensor:

\begin{eqnarray} \label{gdef}
 \hat{g}^{00} &= g^{00} \sqrt{-g} =  &\rho \\
 \hat{g}^{a0} &= g^{a0} \sqrt{-g} =  &\rho v^a \\
 \hat{g}^{ab} &= g^{ab} \sqrt{-g} =  &\rho v^a v^b - \sigma^{ab}
\end{eqnarray}

For Galilean transformations this tensor transforms like a Lorentz
metric.  Instead of searching for a coordinate-dependent Lagrange
function $L(g_{ij}$, $\phi^m,t,x^a)$ we introduce the preferred Galilean
coordinates as independent fields $X^a(x)$, $T(x)$ and try to find a
covariant Lagrange function

\[ L = L(g_{ij},\phi^m,T,X^a). \]

The advantage is that in these variables the conservation laws are
already covariant equations for $X^a$ and T:

\[ \Box T(x) = 0; \; \Box X^a(x) = 0; \]

and for these equations a covariant Lagrange function is well-known:

\[L_0=C_Tg^{ij}T_{,i}T_{,j}+C_X\delta_{ab}g^{ij}X^a_{,i}X^b_{,j}\]

$L_0$ looks really nice in the original Galilean coordinates:

\[ L_0\sqrt{-g}= C_T\rho+ C_X\delta_{ab}(\rho v^av^b-\sigma^{ab})\]

We split the Lagrange function into $L_0$ and a remaining part and
require that the remaining part no longer depends on $X^a$ and $T$:

\[L = L_0 + L_1(g_{ij},\phi^m) \]

This is the simplest way to fulfil our initial requirements.  But
the requirements for $L_1$ are the same as for Lagrange functions in
general relativity with matter fields $\phi^m$.  Thus, we obtain a
one-to-one relation to relativistic theory:

\[L=L_0+L_{rel}(g_{ij},\phi^m)=L_0+R+C_E+L_{matter}(g_{ij},\phi^m)\]

with scalar curvature R, Einstein's cosmological constant $C_E$ and a
matter Lagrangian.  For every relativistic matter Lagrangian we obtain
a related condensed matter theory of some ``ether'' with inner steps
of freedom $\phi^m$.

The non-gravity limit of this theory is de-facto Lorentz ether theory.
Clock time dilation and contraction of rulers may be derived as in
general relativity as a consequence of the symmetry of the matter
Lagrangian.  Thus, the theory solves the conceptual problems of
Lorentz ether theory (insufficient explanation of Lorentz symmetry, no
influence of matter on the ether, no possibility to observe the ether)
by generalization to gravity.  This suggests to name it ``general
ether theory''.

\section{Predictions}

Let's consider now the homogeneous universe solutions of the theory.
Because of the Newtonian background frame, only a flat universe may be
homogeneous. Thus, the theory prefers a flat universe. We make the
ansatz

\[ ds^2 = d\tau^2 - a^2(\tau)(dx^2+dy^2+dz^2). \]

Note that in this ansatz there is no expanding universe but shrinking
rulers.  With some homogeneous matter ($p = k\varepsilon$) we obtain

\begin{eqnarray*}
G^0_0  &=& + C_T /a^6 + 3 C_X /a^2 + C_E +  \varepsilon\\
G^a_a  &=& - C_T /a^6 +   C_X /a^2 + C_E - k\varepsilon\\
\end{eqnarray*}

The parts with $C_T, C_X$ cause effects similar to homogeneously
distributed dark matter with $p=\varepsilon$ resp.
$p=-{1\over3}\varepsilon$.  Thus, the preferred Newtonian frame leads
to observable effects, as a special type of ``dark matter''.

Another prediction is the time-dependence of the cutoff length. The
critical length where the continuous approximation of a condensed
matter theory fails is related with the density:

\[ \rho(x,t) V_{crit} = const \]

In the case of the flat homogeneous universe, the density $\rho(x,t)$
and that's why the cutoff length is approximately constant in harmonic
space coordinates.  That means, the cutoff length is not related with
Planck length, but seems to increase together with the observable
universe expansion.  In the early universe, it was below Planck
length, and in some cosmological future it becomes observable even for
our current devices.

If we set $C_T=C_X=0$, we formally have the Einstein equations of
general relativity in harmonic coordinates.  Nonetheless, the theory
remains different from general relativity in harmonic gauge
(cf. \cite{Kuchar}) even in this limit.  There remains another notion
of completeness of the solution. The solution is complete if it is
defined for all $X^a, T$, the metric should not be complete.  An
example is the black hole collapse.  Starting the collapse with
Minkowski coordinates as initial values for $X^a, T$, we observe that
the part of the solution before horizon formation is already the
complete solution.  The part behind the horizon is unphysical in ether
theory.  Similarly, solutions which do not allow a harmonic time-like
coordinate are forbidden.  Especially, there cannot be solutions with
non-trivial topology or with closed causal loops.

\section{Quantization}

Quantization of a condensed matter theory in a classical Newtonian
framework is unproblematic. The theory solves the problem of
quantization of gravity in the trivial way, by an explicit, physical
regularization.  The preferred Newtonian framework avoids most
conceptual problems of canonical quantization of general relativity
(problem of time \cite{Isham}, topological foam, unitarity and
causality problems \cite{Hartle} and so on), allows to define uniquely
local energy and momentum density for the gravitational field as well
as the Fock space and vacuum state in semi-classical theory.  The
introduction of a preferred frame hypothesis does not require any
modification in special-relativistic QFT.  Hawking radiation occurs
similar to semi-classical GR. The collapsing star evaporates by
Hawking radiation before forming a horizon -- a scenario which has
been discussed for GR too \cite{Gerlach}.

The existence of a preferred frame in our theory allows realistic
causal hidden variable theories like Bohmian mechanics \cite{Bohm},
despite the violation of Bell's inequality \cite{Bell}.  In this
preferred frame, hidden information may be distributed FTL without
causality violation.  In the authors opinion, the EPR principle of
reality \cite{EPR} or causality are much more fundamental principles
compared with observable physical symmetries like local Lorentz
invariance. To reject realism following Bohr \cite{Bohr} or causality
\cite{Price94} is close to an immunization of relativity.  The fact
that both realism and causality may be saved with a preferred frame is
a very strong argument in favour of a preferred frame
\cite{Schmelzer97b}.

\end{document}